\documentclass[]{IEEEtran}
\usepackage{cite,url}
\usepackage{booktabs}
\usepackage{amsmath,amssymb,bm}
\usepackage[dvips]{color}
\usepackage{graphicx}
\usepackage[lined,boxed,commentsnumbered, ruled]{algorithm2e}
\usepackage{booktabs}
\usepackage{subfigure}
\usepackage{amsfonts}
\usepackage{threeparttable}

\newcommand {\bx}{\mathbf{x}}
\newcommand {\by}{\mathbf{y}}
\newcommand {\bzero}{\mathbf{0}}
\newcommand {\pdf}{\mathfrak{g}}

\begin{document}

\title{\fontsize{23pt}{25pt}\selectfont{Uplink Interference Analysis\\ for Two-tier Cellular Networks with Diverse Users\\ under Random Spatial Patterns}}
\author{Wei Bao,~\IEEEmembership{Student Member,~IEEE} and
        Ben Liang,~\IEEEmembership{Senior Member,~IEEE}
\thanks{The authors are affiliated with the Department of Electrical and Computer Engineering, University of Toronto, 10 King's College Road, Toronto, Ontario, Canada (email: \{wbao, liang\}@comm.utoronto.ca). A preliminary version of this work has appeared in IEEE ICCC \cite{OurICCC}.   This work has been supported in part by grants from Bell Canada and the Natural Sciences and Engineering Research Council (NSERC) of Canada. } }
\maketitle
\thispagestyle{empty}
\begin{abstract}
Multi-tier architecture improves the spatial reuse of radio spectrum in cellular networks, but it introduces complicated heterogeneity in the spatial distribution of transmitters, which brings new challenges in interference analysis.  In this work, we present a stochastic geometric model to evaluate the uplink interference in a two-tier network considering multi-type users and base stations. Each type of tier-1 users and tier-2 base stations are modeled as independent homogeneous Poisson point processes, and tier-2 users are modeled as  locally non-homogeneous clustered Poisson point processes centered at tier-2 base stations. By applying a superposition-aggregation-superposition approach, we quantify   the interference at  both tiers.
Our model is also able to capture the impact of two types of exclusion regions, where either tier-2 base stations or tier-2 users are restricted in order to avoid cross-tier interference. As an important application of this analytical model, an intensity planning scenario is investigated, in which we aim to maximize the total income of the network operator with respect to the intensities of tier-2 cells, under constraints on the outage probabilities of tier-1 and tier-2 users. The result of our interference analysis suggests that this maximization can be converted to a standard convex optimization problem.  Finally,
numerical studies further demonstrate the correctness  of our analysis.

\end{abstract}

\section{Introduction} \label{section_intro}
\IEEEPARstart{H}igher capacity, better
service quality, lower power usage, and ubiquitous coverage
are some of the most important objectives in the deployment of wireless cellular networks. To achieve these goals, one efficient approach is to install a second tier of small cells (termed tier-2 cells), such as femtocells, overlapping the original  tier-1 macro cells \cite{Heter_Intro}. Each tier-2 cell is centered at a base station with shorter range and lower cost.  By doing so, the tier-2 network could provide nearby user equipments (UEs) with higher-quality  communication links that require lower power usage.

However, with such tier-2 facilities, interference management becomes more challenging. \emph{First}, the spatial patterns of
 different network components  vary significantly.
Tier-1 BSs are designed and deployed regularly by the network operator;  tier-1 users are randomly distributed in the system; tier-2 BSs are deployed irregularly, sometimes in an ``anywhere plug and play'' manner (e.g., femtocell BSs), implying a high level of spatial randomness; the distribution of tier-2 users are even more complicated: they not only are randomly distributed, but also show spatial correlation, since  they are likely to aggregate around tier-2 BSs.
Because each  network component contributes to the total interference differently, their overall effect is difficult to characterize.
\emph{Second},  tier-2 cells may be classified into different types according to their  communication range (e.g.,  picocell and femtocell) and their local load (e.g., intensity of local UEs),  and  UEs are different in terms of their transmission parameters (e.g., transmission power). Such diverse  tier-2 cells and UEs introduce a more complicated interference environment.  \emph{Third}, in order to alleviate cross-tier interference, a system operator may impose an exclusion region around each tier-1 BS \cite{CDMA_Uplink, SG_ExclusionRegion}, in which either tier-2 BSs or UEs are restricted. This results in a unique pattern of correlation between tiers, bringing additional challenges to accurate interference analysis.

Recent works have applied the theory of stochastic geometry to analyze interference in cellular networks \cite{SG_Totorial3}. Interferers are often modeled as a Poisson point process (PPP), so that the interference created by them  is  the shot noise \cite{SG_Totorial1,SG_Totorial2,SG_Mag} in Euclidean space. The Laplace transform of the shot noise can be derived directly from the Laplace functional \cite{SG_Totorial1,SG_Totorial2} or the generating functional \cite{SG_book1} of the PPP. In this way, the interference can be analyzed mathematically. System metrics, such as outage probability and system throughput can then be
deduced from the Laplace transform.
Employing this  approach, the downlink interference  of multi-tier cellular networks was characterized in \cite{SG_MultiTier,Hetero_downlink2}, and the uplink  interference  of single-tier cellular network was studied in \cite{CDMA_Spatial,CDMA_Spatial2,arXiv_uplink,CDMA_1,CDMA_SameModel}.

However, to analyze the uplink interference in two-tier networks is more challenging, as we need to account for the spatial randomness and correlation of tier-2 UEs aggregating around tier-2 BSs.
Innovative efforts have been made in previous works, but as detailed in Section \ref{section_related}, they only partially resolve the challenges. For example,
  \cite{CDMA_Uplink} evaluated the uplink performance of
two-tier networks based on several levels of approximations. \cite{SG_ICASSP} studied both uplink and downlink interference of two-tier
networks assuming that the UEs transmit at the same power without power control.
Both \cite{CDMA_Uplink} and \cite{SG_ICASSP} considered homogeneous  UEs and tier-2 cells.

In this work, we propose an accurate uplink interference model of two-tier cellular networks, considering multiple types of tier-1 UEs, tier-2 BSs, and tier-2 UEs.  At tier-1, the interference is studied as shot noise corresponding to PPPs.
 At tier-2, we develop a \emph{superposition-aggregation-superposition (SAS)} approach to overcome the challenges in analysis.
  In particular, we show that the interference from all UEs in each tier-2 cell can be equivalently aggregated as a single-point interference source. Through the SAS approach,  we precisely compute the interference of both tiers, avoiding any approximation. 

Furthermore, in order to alleviate cross-tier uplink interference, it is commonly proposed to impose an exclusion region around each tier-1 BS \cite{CDMA_Uplink, SG_ExclusionRegion}, in which tier-2 BSs or tier-2 UEs are restricted. In this paper, we exam the effects of two types of exclusion regions: 1) no tier-2 BSs are allowed within the exclusion regions (BS exclusion);  2) no tier-2 UEs are allowed within the exclusion regions (UE exclusion). Our analytical and simulation observations demonstrate that
 using exclusion regions only bring slightly improvement on the outage performance at a tier-1 BS, but UE-exclusion regions are more effective than BS-exclusion regions with the same exclusion radius.

Another important contribution of this paper is to provide new insights on system design. Through our SAS approach, we show that the coverage probability at tier-1 and tier-2 BSs can be expressed as a product of negative exponential functions of the intensity of tier-2 cells. As an application example of this property, we present an intensity planning scenario, in which we aim to maximize the total income of the network operator with respect to the intensities of tier-2 cells, under constraints on the required outage probabilities of tier-1 and tier-2 UEs. We demonstrate how our analysis can provide an efficient solution to this optimization problem.

The rest of the paper is organized as follows. In section \ref{section_related}, we discuss the relation between our work and prior works.
In Section \ref{section_model}, we present the system model. In Section \ref{section_tier1}, we analyze the interference at tier-1 cells.
 In Section \ref{section_tier2}, we analyze the interference at tier-2 cells. In Section \ref{section_case}, we conduct case studies based on the interference analysis, and the intensity planning problem is presented.
In Section \ref{section_numerical}, we validate our analysis with simulation results. Finally, concluding remarks are given in Section \ref{section_conclusion}.

\section{Related Works}\label{section_related}

For two-tier networks, the downlink interference was well studied through stochastic geometric approaches. For example, Dhillon \textit{et al.}~in \cite{SG_MultiTier} analyzed the downlink outage performance of a heterogeneous network with multiple tiers when the minimum required  signal to interference plus noise ratio (SINR) threshold is greater than $1$. Kim \textit{et al.}~in \cite{Hetero_downlink2}  studied the maximum tier-1 UE and tier-2 cell densities with downlink outage performance constraints.
Singh \textit{et al.}~in \cite{Hetero_downlink3}
 studied the downlink interference with flexible user association.

The analysis of \emph{uplink} interference in two-tier networks is more challenging compared with the downlink case, as we need to account for the spatial randomness and correlation of tier-2 UEs aggregating around tier-2 BSs.
Innovative efforts have been made in previous works. Kishore \textit{et al.}~in
 \cite{SG_MultiTier_nonSG1} studied the uplink performance of a single tier-1 cell and a single tier-2 cell, while and the same authors in \cite{SG_MultiTier_nonSG2} extended it to the case of multiple tier-1 cells and multiple tier-2 cells. However, their models were based on a fixed number of tier-1 and tier-2 cells, without considering the random spatial patterns of UEs and BSs.
  Chandrasekhar and Andrews in \cite{CDMA_Uplink} evaluated the uplink performance of two-tier networks with random  UEs and tier-2 cells. However, several interference components were analyzed based on approximations: 1) the inter-interference of tier-1 cell
 was estimated as a truncated Gaussian random variable; 2) the radius of tier-2 cells was regarded as zero when viewed from the outside; 3) tier-2 UEs were assumed to transmit at the maximum power at the edge of tier-2 cells; and 4) the cross interference from tier-1 UEs to tier-2 BSs only
accounted for the interference from a reference tier-1 cell. Cheung \textit{et al.}~in \cite{SG_ICASSP} studied both uplink and downlink interference of two-tier network based on a Neyman-Scott Process \cite{SG_book1,SG_Cluster1}. However \cite{SG_ICASSP} was also limited in two aspects: 1) all UEs were assumed to transmit at the same power; and 2) tier-2 UEs were assumed to be uniformly distributed in an infinitesimally thin ring around the tier-2 BS. In addition, neither \cite{CDMA_Uplink} nor \cite{SG_ICASSP} considered multi-type  UEs or tier-2 cells. In contrast, our work does not require any of the above approximations, and we further consider multiple types of UEs and tier-2 BSs, and two types of tier-2 exclusion regions.

\section{System Model}\label{section_model}

\subsection{System Topology}
\begin{figure}[tbp]
\centering  \vspace*{0pt}
\includegraphics[scale=.5]{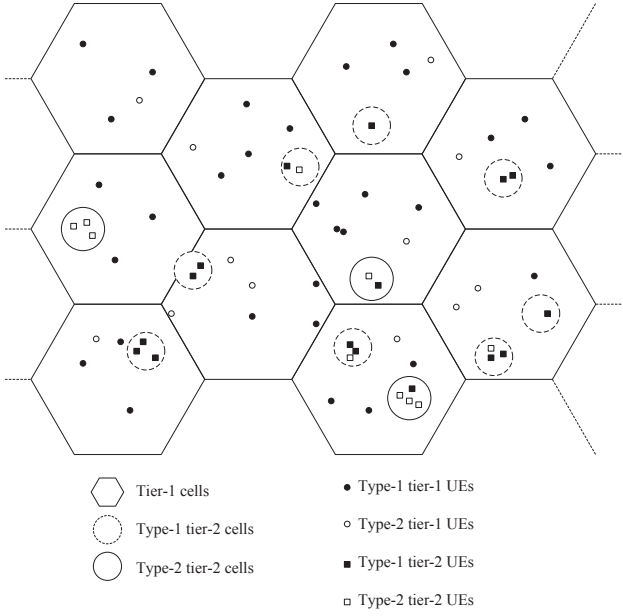}
\caption{System model.}
\label{Figmodel}
\end{figure}

An example of the system topology considered in this work is illustrated in Fig.~\ref{Figmodel}.
We use the terms ``tier-1'' and ``tier-2'' throughout this paper, which are synonymous with ``macro tier'' and ``small-cell tier'' respectively.
First, 
following a common convention in the literature, we assume that the tier-1 cells  form an infinite hexagonal grid on the two-dimensional Euclidean space $\mathbb{R}^2$.
Tier-1 BSs are located at the centers of the hexagons $\mathbb{B}=\{(\frac{3}{2}aR_c, \frac{\sqrt{3}}{2}aR_c+\sqrt{3}bR_c)|a,b \in\mathbb{Z}\}$, where $R_c$  is the radius of the hexagon. Tier-1 UEs are randomly distributed in the system, which are  modeled as PPPs. We assume that there are $M$ types of tier-1 UEs, defined by their different required received power levels. Each type independently forms a homogeneous PPP. Let $\Phi_i$ denote the PPP corresponding to type-$i$ tier-1  UEs. Its intensity is $\lambda_i$.

We consider $N$ types of tier-2 BSs and $K$ types of tier-2 UEs. Different tier-2 BS types are defined in terms of their communication ranges and their local UE densities; different tier-2 UE types are defined in terms of their required received power levels.
Because tier-2 BSs generally have high spatial randomness,  we assume each type of tier-2 BSs  form a homogeneous PPP. Let $\Theta_i$ denote the PPP corresponding to type-$i$ tier-2 BS. Its intensity is $\mu_i$. Each tier-2 BS connects with the Internet core via wired links, which has no influence on the interference analysis.

Each tier-2 BS communicates with different types of local tier-2 UEs surrounding it, composing a tier-2 cell.
Let $R_i$ be the communication radius of each type-$i$ tier-2 BS, with its corresponding tier-2 UEs located within $R_i$ from it.
 \emph{Given} the location of a type-$i$ tier-2 BS at $\bx_0$, we assume that each type of  local tier-2 UEs are independently distributed as a non-homogenous PPP in the disk  centered at $\bx_0$ with radius $R_i$.  Let $\Psi_{ij}(\bx_0)$ denote the PPP of  type-$j$ tier-2 UEs around a type-$i$ tier-2 BS at $\bx_0$. Its intensity at $\bx$ is described by $\nu_{ij}(\bx-\bx_0)$, a non-negative function of the \textit{vector} $\bx-\bx_0$. Note that  the UE intensity   $\nu_{ij}(\bx-\bx_0)=0$ if $|\bx-\bx_0|>R_{i}$. We assume  the tier-2 UEs in one tier-2 cell are also independent with tier-2 UEs in other tier-2 cells as well as tier-1 UEs.  To better understand the distribution of tier-2 BSs and tier-2 UEs, $\Theta_i$   can be regarded as a parent point process on the plane, while $\Psi_{ij}(\cdot)$ is a daughter process associated with a point in the parent point process. Note that the aggregating of tier-2 UEs around tier-2 BSs implicitly defines the location correlation among tier-2 UEs.  In particular, we emphasize that due to the randomness in the location of tier-2 cells, \textit{the locations of different types of tier-2 UEs are dependent and non-Poisson.}

In this paper, we focus on  the closed access scenario. Local tier-2 UEs
only connect to their serving tier-2 BSs, and tier-1 UEs only connect to tier-1 BSs. They are not
transferred to the other tier if they are closer to a BS in that tier. It is also possible that two tier-2 cells are overlapping with each other. In this case, tier-2 UEs maintain  connection to their own serving tier-2 BSs.

Our analytical model requires Poisson assumptions on tier-1 UEs and tier-2 BSs. In Sections \ref{subsection_numerical_correlated} and \ref{subsection_numerical_correlated2}, we present simulation data to study the impact of correlated tier-1 UEs  and tier-2 BSs on analytical accuracy in general and on system performance in the intensity planning problem.


%

\subsection{Pathloss Model and Power Control}
Let $P_t(\bx)$ denote the transmission power at $\bx$ and $P_r(\by)$ denote the received power at $\by$. We assume that $P_r(\by)=\frac{P_t(\bx)g_{\bx}g_{\bx,\by}h_{\bx,\by}}{A|\bx-\by|^{\gamma}}$, where $A|\bx-\by|^{\gamma}$ is the propagation loss function with predetermined constants $A$ and $\gamma$, $g_{\bx}g_{\bx,\by}$ is the shadowing term, which is composed of the near field factor $g_{\bx}$ and far field factor $g_{\bx,\by}$ \cite{CDMA_0,CDMA_1}, and $h_{\bx,\by}$ is the fast fading term. Here, $g_{\bx}$ and $g_{\bx,\by}$ are independently log-normally distributed with given parameters, and $h_{\bx,\by}$ is independently exponentially distributed with unit mean (Rayleigh fading with power normalization).

We follow the conventional assumption that uplink  power control adjusts for propagation losses and shadowing \cite{CDMA_Spatial, CDMA_Uplink, CDMA_0 ,CDMA_1, CDMA_SameModel,CDMA_OpenClose}. The targeted power level for type-$i$ tier-1 UEs  is $P_i$,\footnote{By doing so, we capture the fact that the targeted received powers of different UEs may be different in reality. This provides a more general analysis model than previous works, such as \cite{CDMA_Uplink,CDMA_Spatial,CDMA_1,CDMA_SameModel}, where the targeted received power level is assumed to be a constant.} and the targeted power for  type-$j$  tier-2 UEs in  type-$i$   tier-2 cells is $Q_{ij}$.
Given the targeted received power $P$ (i.e., $P=P_i$ or $P=Q_{ij}$) at $\by$  and transmitter at $\bx$, the transmission power is $\frac{P A|\bx-\by|^{\gamma}}{g_{\bx}g_{\bx,\by}}$. Then, the resultant contribution to interference at $\by'\neq \by$ is
$\frac{P |\bx-\by|^{\gamma}g_{\bx,\by'}h_{\bx,\by'}}{|\bx-\by'|^{\gamma}g_{\bx,\by}}$.

Note that $g_{\bx,\by'}/g_{\bx,\by}$ is still log-normally distributed and is i.i.d.~with respect to different $\bx$, and $h_{\bx,\by}$ is i.i.d.~with respect to different $\bx$ and $\by$.  Let $\mathfrak{g}(\cdot)$ be the probability density function (pdf) of $g_{\bx,\by'}/g_{\bx,\by}$ (log-normal).

In addition, we assume the system is interference limited, such that noise is negligible.


\subsection{Tier-2 Exclusion Regions} \label{subsection_exregion}
\begin{figure}[tbp]
\centering  \vspace*{0pt}
\includegraphics[scale=.65]{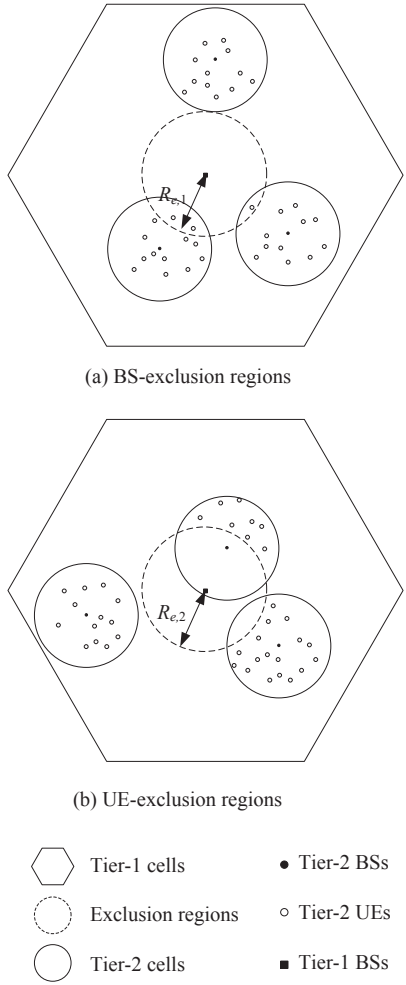}
\caption{Two types of exclusion regions.}
\label{figure_model2}
\end{figure}

To reduce the interference from  tier-2 UEs to tier-1 BSs, exclusion regions of tier-2 cells were proposed in \cite{SG_ExclusionRegion}.  In this paper, we also consider two types of exclusion regions. For \textit{BS exclusion},  we assume that no tier-2 BSs are allowed to locate within $R_{e,1}$ distance from a tier-1 BS, as shown in Fig.~\ref{figure_model2}(a). For \textit{UE exclusion}, we assume that no tier-2 UEs are allowed to locate within $R_{e,2}$ distance from a tier-1 BS, as shown in Fig.~\ref{figure_model2}(b). Let $\mathcal{B}(\bx,R)$ denote the disk region centered at $\bx$ with radius $R$.
  The collection of BS-exclusion and UE-exclusion  regions are denoted by $\mathcal{F}_1=\bigcup_{\bx\in \mathbb{B}} \mathcal{B}(\bx, R_{e,1})$ and $\mathcal{F}_2=\bigcup_{\bx\in \mathbb{B}} \mathcal{B}(\bx, R_{e,2})$, respectively. Thus, with the BS-exclusion regions, the intensity of $\Theta_i$ becomes $0$ in $\mathcal{F}_1$; with the UE-exclusion regions, the intensity of $\Psi_{ij}$ becomes  $0$ in $\mathcal{F}_2$.

   We assume that tier-2 BSs are not protected by exclusion regions, since tier-1 macrocell BSs and UEs are the entrenched equipment whose behavior is difficult to change. Open access has been recognized as an efficient approach to reduce the cross-tier interference from tier-1 UEs to tier-2 BSs. However, it will introduce additional practical concerns (e.g., signaling overhead and network security) as well as substantial challenges in analytical modeling, which is beyond the scope of this paper. Interested readers are referred to \cite{OurMSWIM} for a discussion comparing the open and closed modes.

\subsection{Uplink Multiple Access} \label{subsection_scope}

In this paper, we assume that all uplink communications
occur on the same channel. This analysis
can be extended to non-orthogonal
multiple access schemes, such as CDMA. For
CDMA, a spreading code is applied to transmit a signal, so
that at the receiver, the SIR is equivalent to being multiplied
by the spreading factor \cite{SG_Totorial2,CDMA_Uplink}.

For systems with orthogonal multiple access schemes (e.g., OFDMA), the frequency and time resources are partitioned into multiple orthogonal resource blocks. Different UEs in the same cell use different resource blocks. In this case, a modified version of our model is employed to  provide close approximations for UEs' performance in Section \ref{subsection_OFDM} and  \ref{subsection_numerical_OFDM}.

Furthermore, since we focus on uplink interference analysis, we
assume that the downlink and uplink of the system are
operated separately in different frequency or time. Hence, the downlink interference has no influence on the interference analysis in this paper.

\begin{table}[h!]
\hspace{-0.4cm}
\renewcommand{\arraystretch}{1.3}
\caption{Selected Definition of Variables}
\label{table}
\small
\begin{tabular}{|c| c |   }
\hline
\bfseries Name  & \bfseries Definition \\
\hline
$\Phi_i$                 & The point process of  type-$i$ tier-1 UEs.      \\
\hline
$\Theta_i$          & The point process of  type-$i$ tier-2 BSs.  \\
\hline
                              & The point process of  type-$j$ tier-2  \\
$\Psi_{ij}(\bx_0)$          & UEs   associating with a  type-$i$ tier-2\\
                             &  BS located at $\bx_0$. \\
\hline
$\lambda_i$          & The intensity of  type-$i$ tier-1 UEs.  \\
\hline
$\mu_i$              & The intensity of  type-$i$ tier-2 BSs.  \\
\hline
                              & Given a  type-$i$ tier-2 BS, the intensity of type \\
$\nu_{ij}(\bx)$              &    $j$ tier-2 UEs associating to it, where $\bx$ is the  \\
                              &  relative coordinate with respect to the tier-2 BS. \\
\hline
$I_{1,in,i}$              &  The  interference from  type-$i$ tier-1 UEs \\
                          & inside the typical tier-1 cell to the typical tier-1 BS.   \\
\hline
$I_{1,out,i}$              & The interference from   type-$i$ tier-1 UEs\\
                                 & outside the typical tier-1 cell to the typical tier-1 BS.    \\
\hline
$I_{1,i}$              & The interference from   type-$i$ tier-1 UEs \\
                       & to the typical tier-1 BS. \\
\hline
$I_{1}$              & The interference from  tier-1 UEs \\
                       & to the typical tier-1 BS. \\
\hline
$\widehat{I}_{2,i,j}$              &  The interference from type-$j$ tier-2 UEs inside \\
                                   &  a single type-$i$ tier-2 cell to the typical tier-1 BS.  \\
\hline
$\widehat{I}_{2,i}$              &   The interference from tier-2 UEs inside a single \\
                                 &    type-$i$ tier-2 cell to the typical tier-1 BS. \\
\hline
$I_{2,i}$                       &   The interference from  tier-2 UEs inside all \\
                                 &    type-$i$ tier-2 cells to the typical tier-1 BS. \\
\hline
$I_{2}$                       &   The interference from all tier-2 UEs to the\\
                              & typical tier-1 BS. \\
\hline
$I_{1,i}'$              & The  interference from type-$i$ tier-1 UEs \\
                       & to the typical tier-2 BS. \\
\hline
$I_{1}'$              & The  interference from  tier-1 UEs \\
                       & to the typical tier-2 BS. \\
\hline
$\widehat{I}_{2,i,j}'$              &  The interference from type-$j$ tier-2 UEs inside \\
                                   &  a single type-$i$ tier-2 cell to the typical tier-2 BS.  \\
\hline
$\widehat{I}_{2,i}'$              &   The interference from  tier-2 UEs inside a single \\
                                 &    type-$i$ tier-2 cell to the typical tier-2 BS. \\
\hline
$I_{2,i}'$                       &   The inter-cell interference from tier-2 UEs inside  \\
                                 &    type-$i$ tier-2 cells to the typical tier-2 BS. \\
\hline
$I_{2}'$                       &   The inter-cell interference from  tier-2 UEs \\
                              & to the typical tier-2 BS. \\
\hline
$I_{3,j}'$                       &   The intra-cell interference from type-$j$\\
                              &  tier-2 UEs to the typical tier-2 BS. \\
\hline
$I_{3}'$                       &   The intra-cell interference from tier-2 UEs \\
                              &  to the typical tier-2 BS. \\
\hline
\end{tabular}
\end{table}

\section{Interference to Tier-1 BSs} \label{section_tier1}
In this section, we analyze the uplink interference at tier-1 BSs. Given a reference type-$k$ tier-1 UE, termed the \textit{typical tier-1 UE}, communicating with its BS, termed the \textit{typical tier-1 BS}, we compute the interference from all other tier-1 and tier-2 UEs to the typical BS. The tier-1 cell corresponding to the typical tier-1 BS is denoted as the \textit{typical tier-1 cell}.
Due to stationarity of point processes corresponding to tier-1 UEs, tier-2 BSs, and tier-2 UEs, throughout this section  we will re-define the coordinates so that the typical tier-1 BS is located at $\bzero$. Let $\mathcal{H}(\bx)$ denote the hexagon region centered at $\bx$ with radius $R_c$. Correspondingly, the typical UE is located at some $\bx_U$ that is uniformly distributed in $\mathcal{H}(\bzero)$.
Let $\Phi_{k}^{!}$ denote the point process of all other type-$k$ tier-1 UEs conditioned on the typical UE (i.e., the \textit{reduced Palm point process} with respect to $\Phi_k$).  Since the reduced Palm point process of a PPP has the same distribution as the original PPP,  $\Phi_{k}^{!}$  is still a PPP with intensity $\lambda_k$ \cite{SG_Totorial1}. For presentation convenience, we define $\widetilde{\Phi}_i$, such that $\widetilde{\Phi}_i=\Phi_i$ if $i\neq k$ and $\widetilde{\Phi}_i=\Phi_{i}^{!}$  if $i= k$. Note that the above
typicality definition and the coordination translation follow standard stochastic geometric techniques.

In the following, instead of directly computing the distribution of  interference, we study its Laplace transform (i.e., moment generating function), which fully characterizes its distribution.

\subsection{Interference from Tier-1 UEs to Tier-1 BS}
It is not difficult to compute the Laplace transform of the interference produced by tier-1 UEs to the typical tier-1 BS located at $\bzero$. The following analysis uses a standard stochastic geometric approach and is provided for completeness.

Let $I_{1,in,i}$ denote the total interference from type-$i$ tier-1 UEs inside the typical cell $\mathcal{H}(\bzero)$, and $I_{1,out,i}$ denote the total interference from type-$i$ tier-1 UEs outside the typical cell.  We have
\begin{align}\label{formula_I1in}
I_{1,in,i} =\sum_{\bx\in\widetilde{\Phi}_{i}\bigcap \mathcal{\mathcal{H}}(\bzero)}P_{i}h_{\bx,\bzero},
\end{align}
and
\begin{align}\label{formula_I1out}
\nonumber I_{1,out,i}
=&\sum_{\bx_1\in \mathbb{B}\backslash \{\bzero\}}\sum_{ \substack{\bx\in \widetilde{\Phi}_i \bigcap  \mathcal{H}(\bx_1)}} \frac{P_i |\bx-\bx_1|^{\gamma} g_{\bx,\bzero}h_{\bx,\bzero}}{|\bx|^{\gamma}g_{\bx,\bx_1}},\\
\end{align}
in which $\sum_{ \substack{\bx\in \widetilde{\Phi}_i \bigcap  \mathcal{H}(\bx_1)}} \frac{P_i |\bx-\bx_1|^{\gamma} g_{\bx,\bzero}h_{\bx,\bzero}}{|\bx|^{\gamma}g_{\bx,\bx_1}}$ is the interference from type-$i$ tier-1 UEs in the cell $\mathcal{H}(\bx_1)$, and $\sum_{\bx_1\in \mathbb{B}\backslash \{\bzero\}}\sum_{ \substack{\bx\in \widetilde{\Phi}_i \bigcap  \mathcal{H}(\bx_1)}} \frac{P_i |\bx-\bx_1|^{\gamma} g_{\bx,\bzero}h_{\bx,\bzero}}{|\bx|^{\gamma}g_{\bx,\bx_1}}$ is the overall interference from all type-$i$ tier-1 UEs outside the typical cell.

$I_{1,in,i}$ and $I_{1,out,i}$ can be regarded as shot noises corresponding to $\widetilde{\Phi}_i$. Their Laplace transforms can be derived through Laplace functionals corresponding to PPP:
%
\begin{align}
&\mathcal{L}_{I_{1,in,i}}(s)=\exp\left(-(3\sqrt{3}/2)\lambda_i\left(\frac{sP_iR_c^2}{sP_i+1}\right)\right),
\end{align}
and\footnote{In the numerical computation of (\ref{formula_revisionround2_1}) and (\ref{formula_revisionround2_2}), we have truncated the summation terms up to $ |\mathbf{x}_1|\leq 10R_c$.}
\begin{align}
\nonumber&\mathcal{L}_{I_{1,out,i}}(s)=
\exp\Bigg(-\lambda_i\sum_{\bx_1\in\mathbb{B}\backslash\{\bzero\} } \\
\label{formula_revisionround2_1}&\qquad\qquad\int_{\mathcal{H}(\bx_1)}\int_{\mathbb{R}^+}\frac{sP_i|\bx-\bx_1|^{\gamma}g\pdf(g)} {sP_i|\bx-\bx_1|^{\gamma}g+|\bx|^{\gamma}}dg d\bx\Bigg).
\end{align}
Due to the independence of  $I_{1,in,i}$ and $I_{1,out,i}$, as well as the independence of interference among different tiers,
the Laplace transform of the overall interference from all tier-1 UEs to the typical tier-1 BS can be computed as
\begin{align}\label{formula_begin1}
\mathcal{L}_{I_{1}}(s)=\prod_{i=1}^{M}\mathcal{L}_{I_{1,i}}(s)=\prod_{i=1}^{M}\left(\mathcal{L}_{I_{1,in,i}}(s)\mathcal{L}_{I_{1,out,i}}(s)\right).
\end{align}

\subsection{Interference from  Tier-2 UEs to Tier-1 BS}\label{UEtoBS}
It is much more challenging to characterize the interference from tier-2 UEs to the typical tier-1 BS, which is part of  the core contribution of this work. Because tier-2 UEs are correlated as they aggregate around tier-2 BSs,  the interference cannot be analyzed by a traditional  stochastic geometric approach. Instead, we propose the  following \emph{superposition-aggregation-superposition (SAS)}  method, which exactly captures the interference from tier-2 cells.

\textbf{Interference from One Tier-2 Cell (Superposition):}
In the first step, we study the interference from a single type of tier-2 UEs in a single tier-2 cell. Let $\widehat{I}_{2,i}(\bx_0)$ be the interference from a single type-$i$ tier-2 cell, whose BS is at $\bx_0$, to the typical tier-1 BS. $\widehat{I}_{2,i}(\bx_0)=\sum_{j=1}^{K}\widehat{I}_{2,i,j}(\bx_0)$, where $\widehat{I}_{2,i,j}$ is the interference from all type-$j$ tier-2 UEs in the single type-$i$ tier-2 cell.
We have
\begin{align}
\label{formula_I2ij}\widehat{I}_{2,i,j}(\bx_0)=\sum_{\bx \in \Psi_{ij}(\bx_0)}  \frac{Q_{ij} |\bx-\bx_0|^{\gamma} g_{\bx,\bzero}h_{\bx,\bzero}}{|\bx|^{\gamma}g_{\bx,\bx_0}}.
\end{align}
Its Laplace transform can be derived through the Laplace functional corresponding to $\Psi_{ij}(\bx_0)$,
\begin{align}
\nonumber &\mathcal{L}_{\widehat{I}_{2,i,j}(\bx_0)}(s)\\
=&\exp\Bigg(-\int\limits_{\mathcal{B}(\bzero, R_i)}\int\limits_{\mathbb{R}^+}\frac{sQ_{ij}|\bx|^{\gamma}g\pdf(g)}{sQ_{ij}|\bx|^{\gamma}g+|\bx_0+\bx|^{\gamma}}  dg\nu_{ij}(\bx)d\bx\Bigg).
\label{formula_L2IJ}
\end{align}

In this step, because the location of the tier-2 BS $\bx_0$ is \emph{given},
 different types of tier-2 UEs associated with this tier-2 BS are independent. Thus, the Laplace transform of $\widehat{I}_{2,i}(\bx_0)$ can be computed as
\begin{align}\label{formula_LI2i}
&\mathcal{L}_{\widehat{I}_{2,i}(\bx_0)}(s)=\prod_{j=1}^{K}\mathcal{L}_{\widehat{I}_{2,i,j}(\bx_0)}(s).
\end{align}

Note that the expressions for $\mathcal{L}_{\widehat{I}_{2,i,j}(\bx_0)}(s)$ and $\mathcal{L}_{\widehat{I}_{2,i}(\bx_0)}(s)$ in (\ref{formula_L2IJ}) and (\ref{formula_LI2i})
are functions related to \emph{a unique coordinate} $\bx_0$. This provides an important property for our subsequent analysis, that the interfering signals from all UEs in one tier-2 cell can be equivalently regarded as emission from one \emph{aggregation} point at $\bx_0$. As a consequence, we can use a function of the aggregation point to represent the overall interference from one tier-2 cell.\footnote{Note that for two tier-2 cells (no matter whether or not they overlap), the interfering signals can be equivalently regarded as emission from two points. In this way, (\ref{formula_L2IJ}) and (\ref{formula_LI2i}) accommodate the potential overlapping of two tier-2 cells.}

\textbf{Overall Interference from One Type of Tier-2 Cells (Aggregation):}
Based on the above conclusion, we can study the overall interference from a single type of tier-2 cells.

Let $I_{2,i}$ denote the total interference from type-$i$ tier-2 cells to the typical tier-1 BS,
\begin{align}\label{formula_I2i}
I_{2,i}=\sum_{\bx_0\in\Theta_{i}}\widehat{I}_{2,i}(\bx_0).
\end{align}
Thus, we can derive the Laplace transform of $I_{2,i}$ as follows
\begin{align}
\nonumber &\mathcal{L}_{I_{2,i}}(s)=\mathbf{E}\left(\prod_{\bx_0\in\Theta_{i}}e^{-s\widehat{I}_{2,i}(\bx_0)} \right)\\
=&\mathbf{E}\left(\mathbf{E}\left(\prod_{\bx_0\in\Theta_{i}}e^{-s\widehat{I}_{2,i}(\bx_0)}\bigg|\Theta_{i}\right) \right)\\
\label{formula_generating0}=&\mathbf{E}\left(\prod_{\bx_0\in\Theta_{i}}\mathbf{E}\left(e^{-s\widehat{I}_{2,i}(\bx_0)}\big|\Theta_{i}\right) \right)\\
\label{formula_generating01} =&\mathbf{E}\left(\prod_{\bx_0\in\Theta_{i}}\mathcal{L}_{\widehat{I}_{2,i}(\bx_0)}(s)\right) \\
\label{formula_generating1}=&\exp{\left(-\mu_i\int_{\mathbb{R}^2}(1-\mathcal{L}_{\widehat{I}_{2,i}(\bx_0)}(s))d\bx_0\right)},
\end{align}
where (\ref{formula_generating0}) holds because given $\Theta_i$, the interference from each type-$i$ tier-2 cell is independent with each other; (\ref{formula_generating1}) is obtained since (\ref{formula_generating01}) is in exactly the same form as the generating functional of the PPP $\Theta_i$~\cite{SG_book1}.

\textbf{Overall Interference (Superposition):}
Let $I_2$ denote the total interference from tier-2 UEs to the typical tier-1 BS.
Because multiple types of tier-2 BSs can be regarded as \emph{independent superposition} of each type of tier-2 BSs,
 the Laplace transform  $\mathcal{L}_{I_{2}}$ can be computed as
\begin{align}\label{formula_LI2}
\mathcal{L}_{I_{2}}(s)=\prod_{i=1}^{N}\mathcal{L}_{I_{2,i}}(s).
\end{align}

\subsection{Overall Interference and Outage at Tier-1 Cell}

Since tier-1 UEs and tier-2 UEs are independent, the Laplace transform of the total interference is
\begin{eqnarray}\label{formula_LI}
\mathcal{L}_{I}(s)=\mathcal{L}_{I_1}(s)\mathcal{L}_{I_2}(s).
\end{eqnarray}

Note that the uplink interference occurs at tier-1 BSs and hence its statistics is irrelevant  to  the type of the typical tier-1 UEs communicating with the typical BS.
Then, given an SIR threshold $T$, the outage probability for any type-$k$ tier-1 UE is given by\footnote{The coverage (resp. outage) probability is defined as the probability that SIR is larger (resp. less) than $T$. $\mathbb{P}_{cover,k}=1-\mathbb{P}_{out,k}$.}
\begin{eqnarray}\label{formula_Pout}
\nonumber \mathbb{P}_{out,k}&=&\mathbf{P}(P_k h_{\bx_U,\bzero} < TI)\\
&=&1-\mathcal{L}_I(T/P_k).
\end{eqnarray}

\subsection{Exclusion Region}
In this subsection, we discuss the effect of two types of exclusion regions on the uplink interference analysis at tier-1 BSs.
Considering tier-2 UE-exclusion regions, (\ref{formula_L2IJ}) is affected and  becomes
\begin{align}
\nonumber &\mathcal{L}_{\widehat{I}_{2,i,j}(\bx_0)}(s)=\exp\Bigg(-\int\limits_{\mathcal{B}(\bzero, R_i)}\int\limits_{\mathbb{R}^+}\\
&\frac{sQ_{ij}|\bx|^{\gamma}g\mathbf{1}(\bx+\bx_0\notin \mathcal{F}_2)\pdf(g)}{sQ_{ij}|\bx|^{\gamma}g+|\bx_0+\bx|^{\gamma}}  dg\nu_{ij}(\bx)d\bx\Bigg)\label{formula_exclusion1_1}.
\end{align}

Considering tier-2 BS-exclusion regions, (\ref{formula_generating1}) is affected and $\mathcal{L}_{I_{2,i}}(s)$ becomes
\begin{align}
\mathcal{L}_{I_{2,i}}(s)=\exp{\left(-\mu_i\int_{\mathbb{R}^2\backslash \mathcal{F}_1}(1-\mathcal{L}_{\widehat{I}_{2,i}(\bx_0)}(s))d\bx_0\right)}.
\end{align}

All the other steps in Section \ref{UEtoBS} remain the same.

\section{Interference to Tier-2 BSs} \label{section_tier2}
In this section, we analyze the uplink interference at a reference type-$l$ tier-2 BS, termed the \textit{typical tier-2 BS}, when it is communicating with a reference type-$k$ tier-2 UE, termed the \textit{typical tier-2 UE}. The \textit{typical tier-1 BS} (located at some $\bx_B$) in this section is defined as the tier-1 BS nearest to the typical tier-2 BS.
 Throughout this section we will re-define the coordinates so that the typical tier-2 BS is located at $\bzero$.  Accordingly, $\bx_B$ is uniformly distributed in $\mathcal{H}(\bzero)$. The coordinates of all tier-1 BSs are re-defined as $\mathbb{B}(\bx_B)=\{(\frac{3}{2}aR_c, \frac{\sqrt{3}}{2}aR_c+\sqrt{3}bR_c)+\bx_B|a,b \in\mathbb{Z}\}$. Note that the coordinates in Sections \ref{section_tier1} and \ref{section_tier2} are labeled differently.

\subsection{Interference from Tier-1 UEs to Tier-2 BS}
First, the interference from type-$i$ tier-1 UEs to the typical tier-2 BS is
\begin{align}\label{formula_I1IP}
I'_{1,i}(\bx_B)=&\sum_{\bx_1\in \mathbb{B}(\bx_B)}\sum_{\bx\in\Phi_i\bigcap  \mathcal{H}(\bx_1)} \frac{P_i |\bx-\bx_1|^{\gamma} g_{\bx,\bzero}h_{\bx,\bzero}}{|\bx|^{\gamma}g_{\bx,\bx_1}},
\end{align}
with Laplace transform
\begin{align}
\nonumber &\mathcal{L}_{I'_{1,i}(\bx_B)}(s)=\exp\Bigg(-\lambda_i\sum\limits_{\bx_1\in\mathbb{B}(\bx_B) }\\
\label{formula_revisionround2_2}&\qquad \int_{\mathcal{H}(\bx_1)} \int_{\mathbb{R}^+} \frac{sP_i|\bx-\bx_1|^{\gamma}g\pdf(g)}{sP_i|\bx-\bx_1|^{\gamma}g+|\bx|^{\gamma}} dg d\bx\Bigg).
\end{align}

Due to the independence of different types of tier-1 UEs, the Laplace transform of the interference from all tier-1 UEs is
\begin{align}\label{formula_begin2}
\mathcal{L}_{I'_{1}(\bx_B)}(s)=\prod_{i=1}^{M}\mathcal{L}_{I'_{1,i}(\bx_B)}(s).
\end{align}

\subsection{Inter-Cell Interference from Tier-2 UEs to Tier-2 BS}\label{subsection_tier2UEtoBS}
Similar to the approach in Section \ref{UEtoBS}, we can apply the SAS method to analyze the inter-cell interference from tier-2 UEs to the typical tier-2 BS (excluding co-cell UEs connected with the typical tier-2 BS).

Conditioned on the typical tier-2 BS at $\bzero$, let $\Theta_{l}^{!}$ denote the reduced Palm point process of the other type-$l$ tier-2 BSs. Since $\Theta_l$  is a PPP, $\Theta_{l}^{!}$ has the same distribution as $\Theta_l$ \cite{SG_Totorial1}.  For presentation convenience, we define $\widetilde{\Theta}_i$, such that $\widetilde{\Theta}_i=\Theta_i$ if $i\neq l$ and $\widetilde{\Theta}_i=\Theta_{i}^{!}$  if $i= l$.

Let $\widehat{I}_{2,i}'(\bx_0, \bx_B)$ be the interference from a single type-$i$ tier-2 cell whose BS is at $\bx_0$, to the typical tier-2 BS. $\widehat{I}_{2,i}'(\bx_0, \bx_B)=\sum_{j=1}^{K}\widehat{I}_{2,i,j}'(\bx_0, \bx_B)$, where $\widehat{I}_{2,i,j}'$ is the interference from all type-$j$ tier-2 UEs in the single type-$i$ tier-2 cell. We have
\begin{align}\label{formula_I2IJP}
\widehat{I}_{2,i,j}'(\bx_0, \bx_B)=\sum_{\bx \in \Psi_{ij}(\bx_0)}  \frac{Q_{ij} |\bx-\bx_0|^{\gamma} g_{\bx,\bzero}h_{\bx,\bzero}}{|\bx|^{\gamma}g_{\bx,\bx_0}}.
\end{align}
Its Laplace transform is
\begin{align}
\nonumber &\mathcal{L}_{\widehat{I}_{2,i,j}'(\bx_0,\bx_B)}(s)\\
\label{formula_LI2IJP}=&\exp\Bigg(-\int\limits_{\mathcal{B}(\bzero, R_i)}\int\limits_{\mathbb{R}^+}\frac{sQ_{ij}|\bx|^{\gamma}g\pdf(g)}{sQ_{ij}|\bx|^{\gamma}g+|\bx_0+\bx|^{\gamma}}  dg\nu_{ij}(\bx)d\bx\Bigg).
\end{align}

Due to the independence of different types of UEs in a tier-2 cell, the Laplace transform of $\widehat{I}_{2,i}'(\bx_0, \bx_B)$ is
\begin{align}
&\mathcal{L}_{\widehat{I}_{2,i}'(\bx_0,\bx_B)}(s)=\prod_{j=1}^{K}\mathcal{L}_{\widehat{I}_{2,i,j}'(\bx_0,\bx_B)}(s).
\end{align}

Let $I_{2,i}'(\bx_B)$ denote the overall inter-cell interference from type-$i$ tier-2 cells,
\begin{align}
I_{2,i}'(\bx_B)=\sum_{\bx_0\in\widetilde{\Theta}_{i}}\widehat{I}_{2,i}'(\bx_0,\bx_B).
\end{align}
Similar to the derivation of (\ref{formula_generating1}),  we can derive  its Laplace transform as
\begin{align}
\nonumber &\mathcal{L}_{I_{2,i}'(\bx_B)}(s)\\
\nonumber=&\mathbf{E}\left(\prod_{\bx_0\in\widetilde{\Theta}_{i}}\mathbf{E}\left(e^{-s\widehat{I}_{2,i}'(\bx_0,\bx_B)}|\widetilde{\Theta}_{i}\right) \right)\\
\nonumber=&\mathbf{E}\left(\prod_{\bx_0\in\widetilde{\Theta}_{i}}\mathcal{L}_{\widehat{I}_{2,i}'(\bx_0,\bx_B)}(s)\right) \\
\label{formula_generating22}=&\exp{\left(-\mu_i\int_{\mathbb{R}^2}(1-\mathcal{L}_{\widehat{I}_{2,i}'(\bx_0,\bx_B)}(s))d\bx_0\right)}.
\end{align}

Let $I'_{2}(\bx_B)$ denote the total inter-cell interference from tier-2 UEs to the typical tier-2 BS.
Because multiple types of tier-2 BSs can be regarded as \emph{independent superposition} of each type of tier-2 BSs,
 the Laplace transform of $I'_{2}(\bx_B)$  is
\begin{align}\label{formula_I2p}
\mathcal{L}_{I'_{2}(\bx_B)}(s)=\prod_{i=1}^{N}\mathcal{L}_{I'_{2,i}(\bx_B)}(s).
\end{align}

Note that (\ref{formula_I2IJP})-(\ref{formula_I2p}) do not depend on $\bx_B$, thus $\bx_B$ can be omitted in these formulas. However, if we consider the two types of exclusion regions, (\ref{formula_I2IJP})-(\ref{formula_I2p}) are affected by $\bx_B$, and  $\bx_B$ cannot be omitted.

\subsection{Intra-Cell Interference from Tier-2 UEs to Tier-2 BS}\label{subsection_tier2intra}
In this subsection, we consider the interference within the typical tier-2 cell, given the typical type-$k$ tier-2 UE.  Let $\Psi_{lk}^{!}(\bzero)$ denote the reduced Palm point process of the other type-$k$ UEs in the typical tier-2 cell. Since $\Psi_{lk}(\bzero)$ is a PPP, $\Psi_{lk}^{!}(\bzero)$ has the same distribution as $\Psi_{lk}(\bzero)$. For presentation convenience, we define $\widetilde{\Psi}_{lj}(\bzero)$, such that $\widetilde{\Psi}_{lj}(\bzero)=\Psi_{lj}(\bzero)$ if $k\neq j$ and $\widetilde{\Psi}_{lj}(\bzero)=\Psi_{lj}^{!}(\bzero)$  if $k= j$.

The intra-cell interference from type-$j$  tier-2 UEs is

\begin{align}
I'_{3,j}(\bx_B)=\sum_{\bx\in\widetilde{\Psi}_{lj}(\bzero)}Q_{lj} h_{\bx,\bzero},
\end{align}
with Laplace transform
\begin{align}
\label{formula_end30}\mathcal{L}_{I'_{3,j}(\bx_B)}
=\exp\left(-\frac{sQ_{lj}}{sQ_{lj}+1}\int\limits_{\mathcal{B}(\bzero,R_{l})}\nu_{lj}(\bx)d\bx\right),
\end{align}

Thus, the Laplace transform of the overall interference inside the typical tier-2 cell is
\begin{align}\label{formula_end3}
\mathcal{L}_{I'_{3}(\bx_B)}(s)=\prod_{j=1}^{K} \mathcal{L}_{I'_{3,j}(\bx_B)}(s).
\end{align}
Note that (\ref{formula_end30})-(\ref{formula_end3}) do not depend on $\bx_B$, $\bx_B$ can be omitted in these formulas.  However, if we consider two types of exclusion regions (discussed in Section \ref{subsection_exregiontier2}), (\ref{formula_end30})-(\ref{formula_end3}) are affected by $\bx_B$, and  $\bx_B$ cannot be omitted.

\subsection{Overall Interference and Outage at Tier-2 Cell}
Since $I_1'(\bx_B)$, $I_{2}'(\bx_B)$, and $I_{3}'(\bx_B)$ are independent, the Laplace transform of the total interference given $\bx_B$ is
\begin{eqnarray}\label{formula_begin4}
\mathcal{L}_{I'(\bx_B)}(s)=\mathcal{L}_{I'_{1}(\bx_B)}(s)\mathcal{L}_{I'_{2}(\bx_B)}(s)\mathcal{L}_{I'_{3}(\bx_B)}(s).
\end{eqnarray}

Thus, the Laplace transform of the overall  interference unconditioned on $\bx_B$ is
\begin{align}\label{formula_begin5}
\mathcal{L}_{I'}(s)=\frac{\int_{\mathcal{H}(\bzero)}\mathcal{L}_{I'(\bx_B)}(s)d\bx_B}{\left(3\sqrt{3}/2R_c^2\right)}.
\end{align}

Because $\mathcal{L}_{I'_{2}(\bx_B)}(s)$ and $\mathcal{L}_{I'_{3}(\bx_B)}(s)$ do not depend on $\bx_B$,
\begin{align}\label{formula_begin6}
\mathcal{L}_{I'}(s)=\overline{\mathcal{L}}_{I'_{1}}(s)\cdot \mathcal{L}_{I'_{2}}(s)\cdot\mathcal{L}_{I'_{3}}(s),
\end{align}
where $\overline{\mathcal{L}}_{I'_{1}}(s)=\frac{\int_{\mathcal{H}(\bzero)}\mathcal{L}_{I'_{1}(\bx_B)}(s)d\bx_B}{\left(3\sqrt{3}/2R_c^2\right)}$.

Note that if we consider the two types of exclusion regions, $\mathcal{L}_{I'_{2}(\bx_B)}(s)$ and $\mathcal{L}_{I'_{3}(\bx_B)}(s)$ depend of $\bx_B$, (\ref{formula_begin4})-(\ref{formula_begin5}) rather than (\ref{formula_begin6}) should be employed. 

Finally, given the SIR threshold $T$, the outage probability of the typical tier-2 UE (type-$k$ tier-2 UE in the typical type-$l$  tier-2 cell) is given by
\begin{eqnarray}\label{formula_end4}
 \mathbb{P}_{out,lk}'=1-\mathcal{L}_{I'}(T/Q_{lk}).
\end{eqnarray}

\subsection{Effect of Exclusion Regions}\label{subsection_exregiontier2}
In this subsection, we discuss the effect of the two types of exclusion regions defined in Section \ref{subsection_exregion}
 on the uplink interference analysis at tier-2 BSs.\footnote{Note that the collection of BS-exclusion and UE-exclusion  regions are re-defined as $\mathcal{F}_1(\bx_B)=\bigcup_{\bx\in \mathbb{B}(\bx_B)} \mathcal{B}(\bx, R_{e,1})$ and $\mathcal{F}_2(\bx_B)=\bigcup_{\bx\in \mathbb{B}(\bx_B)} \mathcal{B}(\bx, R_{e,2})$, respectively.}
Considering tier-2 UE-exclusion regions, (\ref{formula_LI2IJP}) and (\ref{formula_end30}) are affected and become
\begin{align}
\nonumber &\mathcal{L}_{\widehat{I}_{2,i,j}'(\bx_0,\bx_B)}(s)=\exp\Bigg(-\int\limits_{\mathcal{B}(\bzero, R_i)}\int\limits_{\mathbb{R}^+}\\
&\frac{sQ_{ij}|\bx|^{\gamma}g\mathbf{1}(\bx+\bx_0\notin \mathcal{F}_2(\bx_B))\pdf(g)}{sQ_{ij}|\bx|^{\gamma}g+|\bx_0+\bx|^{\gamma}}  dg\nu_{ij}(\bx)d\bx\Bigg),
\end{align}
and
\begin{align}
\mathcal{L}_{I'_{3,j}(\bx_B)}
=\exp\left(-\frac{sQ_{lj}}{sQ_{lj}+1}\int\limits_{\mathcal{B}(\bzero,R_{l})\backslash \mathcal{F}_2(\bx_B)}\nu_{lj}(\bx)d\bx\right),
\end{align}

Considering tier-2 BS-exclusion regions, (\ref{formula_generating22}) is affected and becomes
\begin{align}
\mathcal{L}_{I_{2,i}'(\bx_B)}(s)
=\exp{\left(-\mu_i\int_{\mathbb{R}^2\backslash\mathcal{F}_1(\bx_B)}(1-\mathcal{L}_{\widehat{I}_{2,i}'(\bx_0,\bx_B)}(s))d\bx_0\right)}.
\end{align}

All the other steps in Section \ref{subsection_tier2UEtoBS} and \ref{subsection_tier2intra} remain the same. Note that by considering the exclusion regions, (\ref{formula_I2IJP})-(\ref{formula_end3})  depend on $\bx_B$. In this case, we should employ (\ref{formula_begin4})-(\ref{formula_begin5}), rather than (\ref{formula_begin6}) to compute the Laplace transform of $I'$.

\section{Case Studies}\label{section_case}
In this section, we present several important case studies based on our analysis in Section \ref{section_tier1} and \ref{section_tier2}. First, we discuss the effects of several network parameters based on a single-type scenario.
Second, we present a modified version of our model, which provides close approximations for UEs' performance in systems using orthogonal multiple access schemes. Third, we present a negative exponential property and the intensity planning problem.

\subsection{Single-Type Scenario}
In this subsection, we present a simple case with only one type  of tier-1 UEs, one type of  tier-2 BSs, and one type of  tier-2 UEs (i.e., $M=N=K=1$). Exclusion regions are not considered.

First, according to (\ref{formula_I1in})-(\ref{formula_Pout}),
the outage probability of tier-1 UEs is $\mathbb{P}_{out}=1-\mathbb{P}_{1,in}\mathbb{P}_{1,out}\mathbb{P}_{2}$,
where
\begin{align}
\label{formula_case11}\mathbb{P}_{1,in}=&\mathcal{L}_{I_{1,in,1}}(T/P_1)=\exp\left(-\lambda_1(3\sqrt{3}/2)\left(\frac{TR_c^2}{T+1}\right)\right),\\
\nonumber\mathbb{P}_{1,out}=&\mathcal{L}_{I_{1,out,1}}(T/P_1)=\exp\Bigg(-\lambda_1\sum_{\bx_1\in\mathbb{B}\backslash\{\bzero\} }\\
\label{formula_case12}&\int_{\mathcal{H}(\bx_1)}\int_{\mathbb{R}^+}\frac{T|\bx-\bx_1|^{\gamma}g\pdf(g)}{T|\bx-\bx_1|^{\gamma}g+|\bx|^{\gamma}} dgd\bx\Bigg),\\
\nonumber\mathbb{P}_{2} =&\mathcal{L}_{I_{2,1}}(T/P_1)=\exp\Bigg(-\mu_1\int_{\mathbb{R}^2}\bigg(1-\exp\bigg(-\int\limits_{\mathcal{B}(\bzero, R_1)}\\
\label{formula_case13}&\int\limits_{\mathbb{R}^+}\frac{\frac{Q_{11}T}{P_1}|\bx|^{\gamma}g\nu_{11}(\bx)\pdf(g)}{\frac{Q_{11}T}{P_1}|\bx|^{\gamma}g+|\bx_0+\bx|^{\gamma}}
dg d\bx\bigg)\bigg)d\bx_0\Bigg).
\end{align}

Then, according to (\ref{formula_I1IP})-(\ref{formula_end4}), the outage probability of tier-2 UEs is $\mathbb{P}_{out}'=1-\mathbb{P}'_{1}\mathbb{P}'_{2}\mathbb{P}'_{3}$, where
\begin{align}
\nonumber \mathbb{P}'_{1}=&\mathcal{L}_{I'_{1}}(T/Q_{11})=\frac{1}{3\sqrt{3}/2R_c^2}\int_{\mathcal{H}(\bzero)}\exp\Bigg(-\lambda_1\sum\limits_{\bx_1\in\mathbb{B}(\bx_B) }\\
\label{formula_case14}&\int_{\mathcal{H}(\bx_1)} \int_{\mathbb{R}^+} \frac{\frac{P_1T}{Q_{11}}|\bx-\bx_1|^{\gamma}g\pdf(g)}{\frac{P_1T}{Q_{11}}|\bx-\bx_1|^{\gamma}g+|\bx|^{\gamma}} dgd\bx\Bigg)d\bx_B,\\
\nonumber \mathbb{P}'_{2}
=&\mathcal{L}_{I'_{2}}(T/Q_{11})=\exp\Bigg(-\mu_1\int_{\mathbb{R}^2}\bigg(1-\exp\bigg(-\int\limits_{\mathcal{B}(\bzero, R_1)}\\
\label{formula_case15}&\qquad\int\limits_{\mathbb{R}^+}\frac{T|\bx|^{\gamma}g\nu_{11}(\bx)\pdf(g)}{T|\bx|^{\gamma}g+|\bx_0+\bx|^{\gamma}}
 dgd\bx\bigg)\bigg)d\bx_0\Bigg),\\
\mathbb{P}'_{3}
\label{formula_case16}=&\mathcal{L}_{I'_{3}}(T/Q_{11})=\exp\left(-\frac{T}{T+1}\int\limits_{\mathcal{B}(\bzero,R_{1})}\nu_{11}(\bx)d\bx\right).
\end{align}

Through (\ref{formula_case11})-(\ref{formula_case16}), we can observe important effects of different network parameters:
\subsubsection{Effects of Targeted Received Power}\label{subsection_optimalP}
$\mathbb{P}_{1,in}=\mathcal{L}_{I_{1,in,1}}(T/P_1)$ corresponds to the coverage probability of tier-1 UEs if there is only co-tier intra-cell interference from tier-1 UEs, which is irrelevant to $P_1$. Even if $P_1$ is changed, both the received signal level and the co-tier interference level at tier-1 BSs are scaled by the same factor, leading to a constant signal to interference ratio. Thus, $\mathbb{P}_{1,in}$ does not change in this case. Similarly, $\mathbb{P}_{1,out}$ corresponds to the coverage probability of tier-1 UEs if there is only co-tier inter-cell interference from tier-1 UEs, which is also irrelevant to $P_1$.

For the same reason, $\mathbb{P}'_{2}$ (resp. $\mathbb{P}'_{3}$) corresponds to the coverage probability of tier-2 UEs if there is only co-tier inter-cell interference (resp. intra-cell interference) from tier-2 UEs, which is  irrelevant to $Q_{11}$.

$\mathbb{P}_{2}$ and $\mathbb{P}'_{1}$ are related to the cross-tier SIR. Increasing $\frac{Q_{11}}{P_1}$ leads to higher  cross-tier SIR at tier-2 BSs, but lower cross-tier SIR at tier-1 BSs. Thus, $\mathbb{P}'_{1}$  is increased and $\mathbb{P}_{2}$ is decreased.

As a example, we may design $\frac{Q_{11}}{P_1}$ to minimize the overall average outage probabilities of both tier-1 and tier-2 UEs, which can be computed as
\begin{align}
\overline{\mathbb{P}}_{out}=\frac{\mathbb{P}_{out}\lambda_1+\mathbb{P}_{out}'\mu_1\int_{\mathcal{B}(\bzero,R_{1})}\nu_{11}(\bx)d\bx }{\lambda_1+\mu_1\int_{\mathcal{B}(\bzero,R_{1})}\nu_{11}(\bx)d\bx}.
\end{align}
Numerical methods can be applied to search for the optimal  $\frac{Q_{11}}{P_1}$ values, which will be further discussed in Section \ref{subsection_numerical_optimalP}.
 Note that  $\mathbb{P}_{2}$ and $\mathbb{P}'_{1}$ are the only parts to be recomputed under different $\frac{Q_{11}}{P_1}$, which reduces the complexity in the numerical search.

\subsubsection{Effects of $\nu_{11}(\bx)$}\label{subsection_differentV}
Let $\overline{\nu}_{11}=\int_{\mathcal{B}(\bzero,R_{1})}\nu_{11}(\bx)d\bx$ indicate the average number of tier-2 UEs in a tier-2 cell. We are interested in studying the outage performance under the same $\overline{\nu}_{11}$, but different $\nu_{11}(\bx)$.

If $\nu_{11}(\bx)$ becomes more dense at locations with larger $|\bx|$ (but less dense at locations with smaller $|\bx|$),  tier-2 UEs are more likely to locate at cell edges where they transmit with higher power levels. As a consequence, the interference from tier-2 UEs is increased and the outage probabilities of both tier-1 and tier-2 UEs are increased. Further numerical studies are given in Sections \ref{subsection_numerical_differentV}.

\subsection{Orthogonal Multiple Access}\label{subsection_OFDM}
It is desirable to study systems using orthogonal multiple access schemes, such as OFDMA. In this case, the frequency
and time resources are partitioned into $n$ orthogonal resource blocks. Each BS randomly allocates one unused
resource block to one UE in the cell.\footnote{If there are more than $n$ UEs in one cell,  we assume that the BS randomly selects $n$ UEs to allocate resource blocks.}
However, this introduces complicated coupling between the point processes of BSs and UEs, which are difficult to characterize based on standard stochastic geometric analysis \cite{arXiv_uplink}.

In this paper, we employ the following modified version of our analysis to approximately  characterize orthogonal multiple access systems:
(a) Intra-cell interference terms are not accounted (i.e., $I_{1,in,i}$ and $I'_{3,j}$ are regarded as zero). (b) Inter-cell interfering UEs are equivalently viewed as
independently thinned point processes with probability $\frac{1}{n}$.  The equivalent intensity of type-$i$ tier-1 UEs is $\frac{\lambda_i}{n}$, and the intensity of type-$j$ tier-2 UEs in a type-$i$ tier-2 cell is characterized by $\frac{\nu_{ij}(\bx-\bx_0)}{n}$. Note that in the orthogonal multiple access mode, the resultant interfering UEs actually correspond to dependently thinned point processes. We use the independently thinned point processes to approximate their corresponding dependently thinned ones.\footnote{This approximation is applicable to the systems where the average number of UEs per cell is less than the number of resource blocks $n$.}

Simulation results in Section \ref{subsection_numerical_OFDM} indicate that the above method leads to closely approximated outage probabilities for both tier-1 and tier-2 UEs.


\subsection{Negative Exponential Property and Intensity Planning}\label{subsection_intensityplan}
\subsubsection{Negative Exponential Property}
Through our discussion in Section \ref{section_tier1} and \ref{section_tier2}, we observe that $\mathcal{L}_{I_2}(s)$  and $\mathcal{L}_{I'_2}(s)$ can be expressed in the form of $\mathcal{L}_{I_2}(s)=\prod_{i=1}^{N}\exp(-\mu_i C_i(s))$ and $\mathcal{L}_{I'_2}(s)=\prod_{i=1}^{N}\exp(-\mu_i C'_i(s))$,
where
 \begin{align}
C_i(s)=&\int_{\mathbb{R}^2}\bigg(1-\mathcal{L}_{\widehat{I}_{2,i}(\bx_0)}(s)\bigg)d\bx_0,\\
C'_i(s)=&\int_{\mathbb{R}^2}\bigg(1-\mathcal{L}_{\widehat{I}'_{2,i}(\bx_0)}(s)\bigg)d\bx_0,
\end{align}
according to (\ref{formula_generating1})-(\ref{formula_LI2}) and (\ref{formula_generating22})-(\ref{formula_I2p}) respectively. The neat form expressions for $\mathcal{L}_{I_2}(s)$  and $\mathcal{L}_{I'_2}(s)$ are referred to as the \emph{negative exponential property} of tier-2 cell intensities. Next, we will present the usefulness of this  negative exponential property in an intensity planning problem.

\subsubsection{Intensity Planning} \label{subsubsection_application}

Consider a network upgrading scenario. The original network is a one-tier network with multi-type  UEs,
which  matches the system model of  the tier-1 network in this paper (discussed in Section \ref{section_model}). All the parameters of the original one-tier network are given, including $R_c, \lambda_i$, and $P_i$. The operator aims  to update the network to  a two-tier network, with diverse tier-2 cells and tier-2 UEs. The tier-2 network also matches the  same system model of  the tier-2 network in this paper. The parameters of each type of tier-2 cells are also given, including  $R_i, Q_{ij}$, and $\nu_{ij}(\mathbf{x})$. Exclusion regions are  not considered.
After upgrading the network, suppose the operator makes extra income $U_i(\mu_i)$ by type-$i$  tier-2 cells,  where $U_i$ is a non-decreasing concave function of the intensity $\mu_i$. Thus, the total income of updating the network is $\sum_{i=1}^N U_i(\mu_i)$. To guarantee the uplink quality of tier-1 and tier-2 UEs, the outage probability of the  type-$j$ tier-1 UEs cannot be greater than $\mathbb{P}_{target,j}$; and the outage probability of the  type-$k$ tier-2 UEs in type-$l$ tier-2 cells cannot be greater than $\mathbb{P}'_{target,lk}$. These outage probability constraints restrict the intensities of tier-2 cells. 
In summary, we can establish a utility maximization problem:
 \begin{align}
  \label{formula_optimization}& \underset{\mu_i, \forall i}{\text{maximize}},
& & \sum_{i=1}^{N}U_i(\mu_i)\\
& \text{subject to}
 \label{formula_constraint_T1}& & \mathbb{P}_{out,j} (\mu_1,\ldots,\mu_N)\leq \mathbb{P}_{ target,j},\quad\forall j,\\
  \label{formula_constraint_T2}& & &\mathbb{P}'_{out,lk} (\mu_1,\ldots,\mu_N)\leq \mathbb{P}'_{ target,lk},\quad\forall l, k,\\
\label{formula_optimal}& & &\mu_i\geq 0,\quad\forall i,
\end{align}
where $\mathbb{P}_{out,j} (\mu_1,\mu_2,\ldots,\mu_N)$ and $\mathbb{P}'_{out,lk} (\mu_1,\ldots,\mu_N)$ are
 the outage probabilities of type-$j$ tier-1 UEs and type-$k$ tier-2 UEs in type-$l$ tier-2 cells
 under $\mu_1,\mu_2,\ldots,\mu_N$, computed as (\ref{formula_Pout}) and (\ref{formula_end4}) respectively.

By the negative exponential property, the outage probability constraints (\ref{formula_constraint_T1}) can be converted to inequalities:
 \begin{align}
\prod_{i=1}^{N}\exp (-A_{ij}\mu_i)&\geq \frac{1-\mathbb{P}_{target,j}}{{\mathcal{L}_{I_1}(T/P_j)}},
\end{align}
which is equivalent to
 \begin{align}\label{formula_constraint2}
\sum_{i=1}^{N}A_{ij}\mu_i\leq B_j,
\end{align}
where $A_{ij}=C_i(T/P_j)$ and $B_{j}=-\log\left(\frac{1-\mathbb{P}_{target,j}}{{\mathcal{L}_{I_1}(T/P_{j})}}\right)$.

Similarity, the outage probability constraints (\ref{formula_constraint_T2}) can be converted to inequalities:
 \begin{align}
\prod_{i=1}^{N}\exp (-A'_{ilk}\mu_i)&\geq \frac{1-\mathbb{P}'_{target,lk}}{{\overline{\mathcal{L}}_{I'_1}(T/Q_{lk})\mathcal{L}_{I'_3}(T/Q_{lk})}},
\end{align}
which is equivalent to
 \begin{align}\label{formula_constraint22}
\sum_{i=1}^{N}A'_{ilk}\mu_i\leq B'_{lk},
\end{align}
where $A'_{ilk}=C'_i(T/Q_{lk})$ and $B'_{lk}=-\log\left(\frac{1-\mathbb{P}'_{target,lk}}{{\overline{\mathcal{L}}_{I'_1}(T/Q_{lk})\mathcal{L}_{I'_3}(T/Q_{lk})}}\right)$.

Note that $A_{ij}$, $B_{j}$, $A'_{ilk}$, and $B'_{lk}$ can be computed with the given network parameters. Thus, the constraints (\ref{formula_constraint2}) and (\ref{formula_constraint22}) represent \emph{linear intensity tradeoff}. As a consequence, the original optimization problem (\ref{formula_optimization})-(\ref{formula_optimal}) is converted to a convex optimization problem with simple linear constraints, which can be solved efficiently.  Further numerical studies are given in Sections \ref{subsection_numerical_intensityplan}.

This application example demonstrates that our analysis can be used to convert the complicated outage probability constraints into simple linear constraints, facilitating tractable problem solutions.

\section{Numerical Study} \label{section_numerical}
In this section, we present a numerical study to validate the accuracy and utility of our analysis model.
Unless otherwise stated, $R_c=1$ km, $\gamma=4$, the shadowing term is log-normal with mean $0$ and standard deviation $4$ dB, fast fading is Rayleigh with unit mean.  Each simulation data point is averaged over $10000$ trials.
Error bars in the figures show the $95\%$ confidence intervals for simulation results. Some plot points are slightly shifted to avoid overlapping error bars for easier inspection.  The network parameters used in all figures are listed in Table \ref{tabel_para}.

\begin{table*}[!t]
\caption{List of Parameters Used in Numerical Studies.}
\label{tabel_para} \centering

\begin{threeparttable}
\begin{tabular}{|c|c||c|c|c|c|c|c|c|}
\hline
\textbf{Section}&\textbf{Figure}& $M,N,K$ & Power (dBm) & $\lambda,\mu$ (units/km$^2$) & Radius (km) & $\nu(\cdot)$ (units/km$^2$)\tnote{a} & $T$& Special  \\
\hline
\textbf{\ref{subsection_numerical_s}}&\textbf{\ref{FigA}(a)-\ref{FigA}(d)}& $1,1,1$ & $(P_1,Q_{11})=(-70,-70)$ & Various & $R_1=0.2$ & $\nu_{11}(\bx)=20$ & $0.1$ or $1$ & None  \\
\hline
\textbf{\ref{subsection_numerical_s}}&\textbf{\ref{FigA}(e)}& $1,1,1$ & $(P_1,Q_{11})=(-70,-70)$ & $(\lambda_1,\mu_1)=(0.5,0.5)$ & $R_1=0.2$ & $\nu_{11}(\bx)=20$ & Various  & None  \\
\hline
&&   & $(P_1,P_{2})=(-67,-65.2)$ &  &  & $\nu_{11}(\bx)=10$ &   &  \\
\textbf{\ref{subsection_numerical_m}}&\textbf{\ref{FigA}(f)}& $2,2,2$ & $(Q_{11},Q_{12})=(-70,-64)$ & $(\lambda_1,\lambda_2)=(0.5,0.5)$ & $R_1=0.1$ & $\nu_{12}(\bx)=15$ & 0.1  & None  \\
&&  & $(Q_{21},Q_{22})=(-67,-67)$ &$(\mu_1,\mu_2)=(1,1)$ &$R_2=0.2$  & $\nu_{21}(\bx)=5$ &  &   \\
&  && & &  & $\nu_{22}(\bx)=20$ &  &   \\
\hline
\textbf{\ref{subsection_numerical_ex}}&\textbf{\ref{FigA}(g)-\ref{FigB}(a)}& $1,1,1$ & $(P_1,Q_{11})=(-70,-70)$ & $(\lambda_1,\mu_1)=(0.5,0.5)$ & $R_1=0.2$ & $\nu_{11}(\bx)=20$ & $0.1$ or $1$  & \scriptsize{Exclusion Regions} \\
\hline
\textbf{\ref{subsection_numerical_optimalP}}&\textbf{\ref{FigB}(b)}& $1,1,1$ & Various & $(\lambda_1,\mu_1)=(0.5,1)$ & $R_1=0.2$ & $\nu_{11}(\bx)=20$ & $0.1$  & None  \\
\hline
\textbf{\ref{subsection_numerical_differentV}}&\textbf{\ref{FigB}(c)-\ref{FigB}(d)}& $1,1,1$ & $(P_1,Q_{11})=(-70,-70)$ & $\lambda_1=0.05$ & $R_1=0.2$ & \scriptsize{None-homogeneous} & $0.1$  & None  \\
\hline
&&   & $(P_1,P_{2})=(-67,-65.2)$ &  &  & $\nu_{11}(\bx)=16$ &   &  \\
\textbf{\ref{subsection_numerical_OFDM}}&\textbf{\ref{FigB}(e)}& $2,2,2$ & $(Q_{11},Q_{12})=(-70,-64)$ & $(\lambda_1,\lambda_2)=(1.6,1.6)$ & $R_1=0.1$ & $\nu_{12}(\bx)=48$ & 1  & \scriptsize{OFDM, $n=16$}  \\
&&  & $(Q_{21},Q_{22})=(-67,-67)$ &$(\mu_1,\mu_2)=(1,1)$ &$R_2=0.2$  & $\nu_{21}(\bx)=32$ &  &   \\
&&  &  & & & $\nu_{22}(\bx)=32$ &  &   \\
\hline
\textbf{\ref{subsection_numerical_intensityplan}}&\textbf{\ref{FigB}(f)}& $2,2,1$ & $(P_1,P_{2})=(-67,-66)$ & $(\lambda_1,\lambda_2)=(0.1,0.1)$ & $R_1=0.2$ & $\nu_{11}(\bx)=10$ & 0.05  & \scriptsize{Intensity Tradeoff}  \\
&&  & $(Q_{11},Q_{12})=(-60,-59.2)$ &$(\mu_1,\mu_2)$ various &$R_2=0.2$  & $\nu_{12}(\bx)=8$ &  &   \\
\hline
&&   & $(P_1,P_{2})=(-67,-65.2)$ &  && $\nu_{11}(\bx)=10$ &   &  \\
\textbf{\ref{subsection_numerical_correlated}}&\textbf{\ref{FigB}(g)-\ref{FigB}(h)}& $2,2,2$ & $(Q_{11},Q_{12})=(-70,-64)$ & $(\lambda_1,\lambda_2)=(0.5,0.5)$ & $R_1=0.1$ & $\nu_{12}(\bx)=15$ & 0.1  & \scriptsize{With correlation}  \\
&&  & $(Q_{21},Q_{22})=(-67,-67)$ &$(\mu_1,\mu_2)=(1,1)$ &$R_2=0.2$  & $\nu_{21}(\bx)=5$ &  &   \\
&&  & & &  & $\nu_{22}(\bx)=20$ &  &   \\
\hline
& &                                            & $(P_1,P_{2})=(-67,-66)$     &                                   &   & $\nu_{11}(\bx)=10$ &   &  \\
\textbf{\ref{subsection_numerical_correlated2}}&\textbf{\ref{FigB}(i)}& $2,2,2$ & $(Q_{11},Q_{12})=(-67,-67)$ & $(\lambda_1,\lambda_2)=(0.2,0.1)$ & $R_1=0.2$ & $\nu_{12}(\bx)=5$ & 0.05  & \scriptsize{With correlation}  \\
&&                                             & $(Q_{21},Q_{22})=(-70,-64)$ &$(\mu_1,\mu_2)$ various             &$R_2=0.2$   & $\nu_{21}(\bx)=5$ &  &   \\
& & & & &                                                                                                                   & $\nu_{22}(\bx)=5$ &  &   \\
\hline
\end{tabular}
\begin{tablenotes}
\item[a] Corresponding to the intensities inside the cell range.
\end{tablenotes}

\end{threeparttable}
\end{table*}

\subsection{Model Comparison}\label{subsection_numerical_s}
First, we compare our system model with the model with approximations in \cite{CDMA_Uplink}.
Since the authors of \cite{CDMA_Uplink} did not consider multi-type UEs or BSs, our comparison is based on the case of single-type tier-1 UEs, tier-2 cells, and tier-2 UEs (i.e., $M=N=K=1$). We use all four approximating assumptions stated in Section \ref{section_related} in the way of \cite{CDMA_Uplink}.

\begin{figure*}
\centering
\vspace*{0pt}
\includegraphics[scale=1]{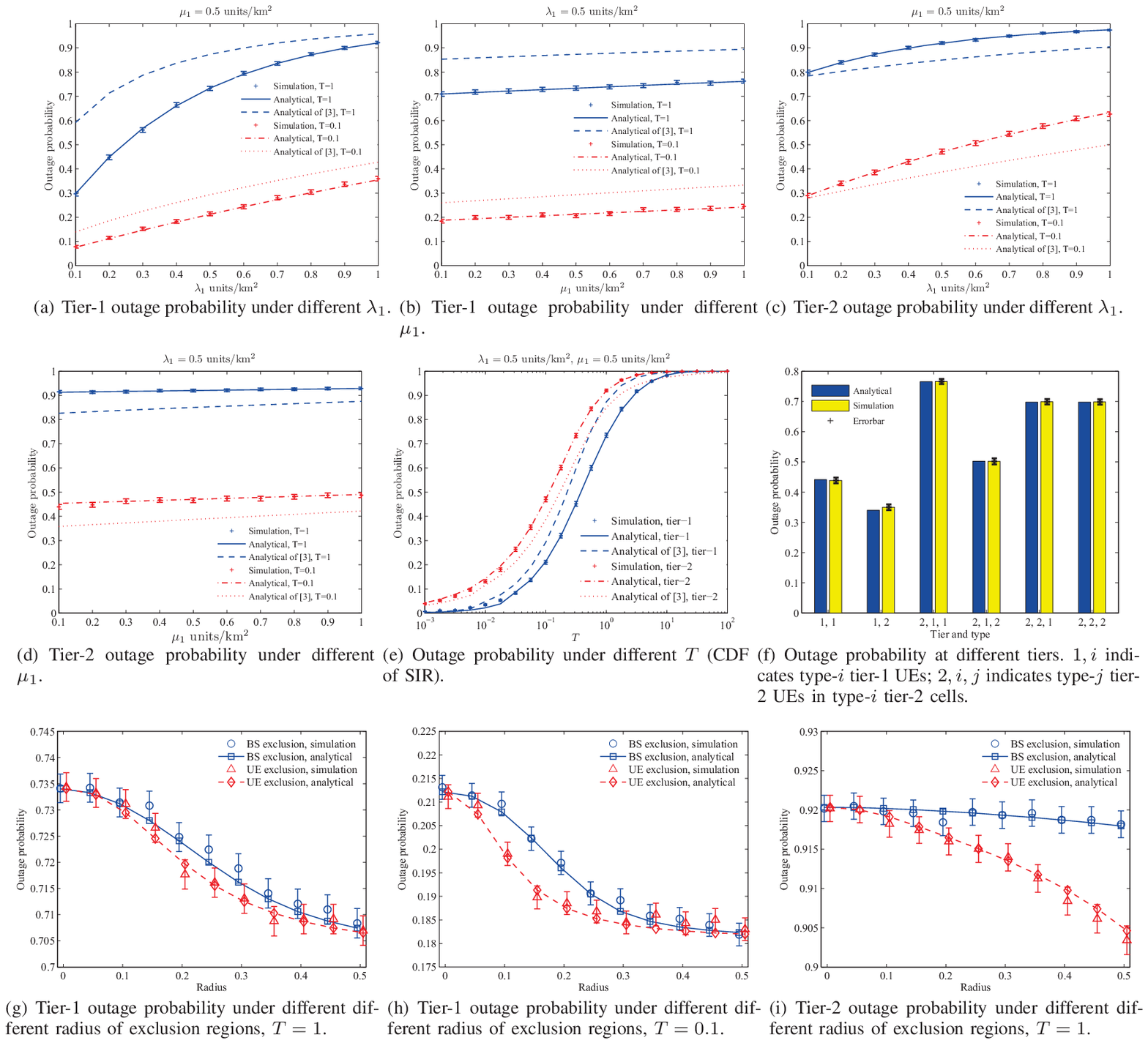}
\vspace{-0.5cm}
\caption{Numerical studies. }
\label{FigA}
\end{figure*}

Figs.~\ref{FigA}(a)-\ref{FigA}(b) show the uplink outage  probability of tier-1 cells under different $\lambda_1$ and $\mu_1$ respectively. 
The figures illustrate that our analytical results are accurate and offer improvement over the model in \cite{CDMA_Uplink}.
In \cite{CDMA_Uplink}, because tier-2 UEs are assumed to be located at the edge of tier-2 cells and transmitting with maximum power,  the interference from tier-2 UEs to tier-1 BSs (as well as tier-2 BSs) is overestimated.  Also, when the tier-1 inter-cell interference is approximated as truncated Gaussian, larger evaluation error occurs.  Overall, the outage probabilities of tier-1 uplinks are overestimated by the model in  \cite{CDMA_Uplink}.

Figs.~\ref{FigA}(c)-\ref{FigA}(d) show the  uplink outage  probability of tier-2 cells under different $\lambda_1$ and $\mu_1$ respectively. The figures again illustrate that our analytical results are accurate and offer improvement. In  \cite{CDMA_Uplink}, because the interference from tier-1 UEs outside the reference tier-1 cell is ignored, the cross-tier interference from tier-1 UEs to tier-2 BSs is greatly underestimated. Even though the co-tier interference from tier-2 UEs to tier-2 BSs is overestimated, that cannot compensate for the underestimation of the cross-tier interference. Hence, overall, the outage probabilities of tier-2 uplinks are underestimated by the model in  \cite{CDMA_Uplink}.

Notably, as we increase $\lambda_1$, more tier-1 interference is ignored, causing larger approximation error;  when we increase $\mu_1$, the overestimation of the co-tier interference cancels more of the underestimation of the cross-tier interference, leading to an overall smaller estimation error.

Fig.~\ref{FigA}(e) shows the  outage  probability under different $T$ (i.e., the CDF of SIR). This figure further confirms that our
analytical results are more accurate than the alternatives.

\begin{figure*}
\centering
\vspace*{0pt}
\includegraphics[scale=1]{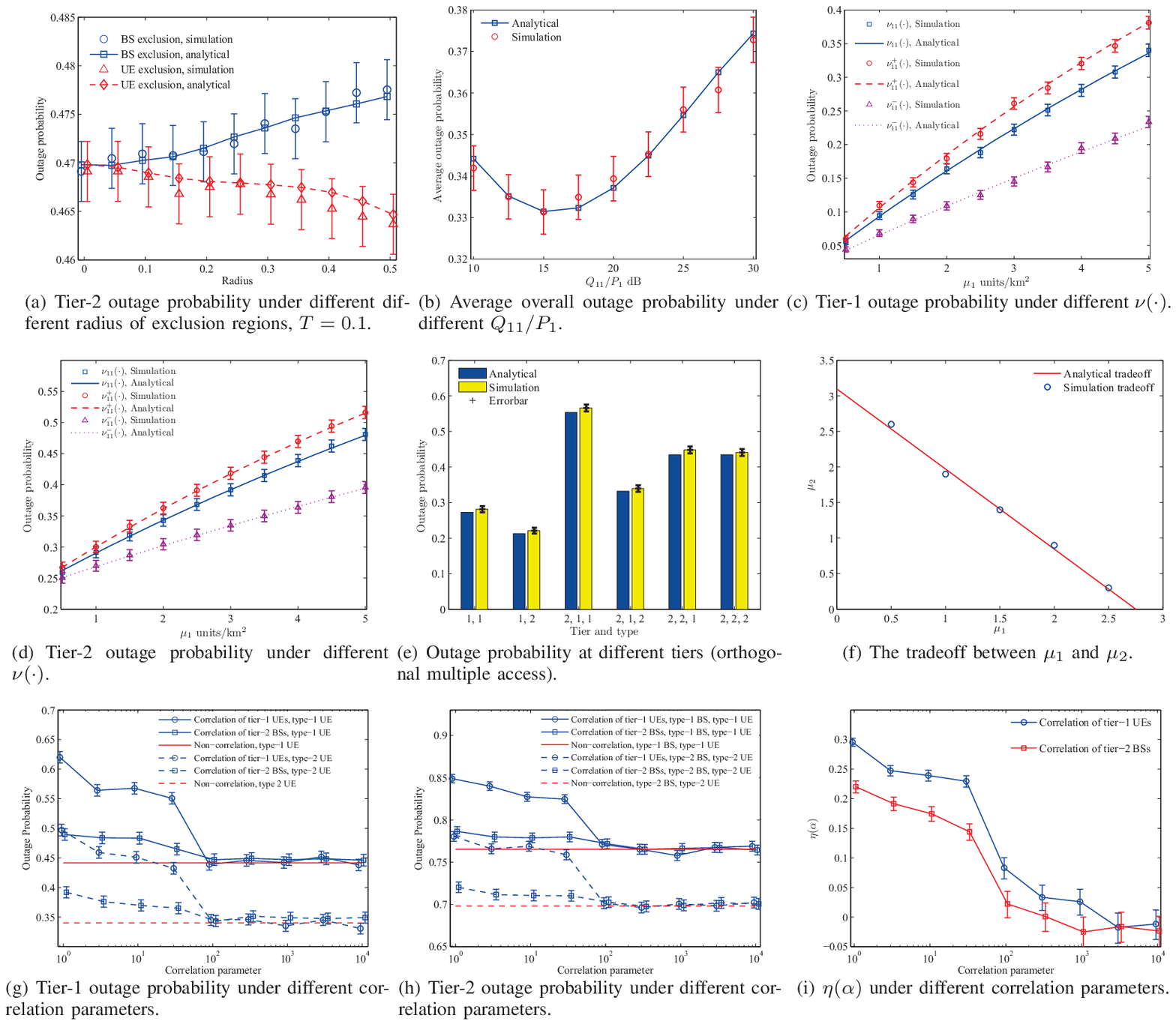}
\vspace{-0.5cm}
\caption{Numerical studies (cont'd). }
\label{FigB}
\end{figure*}

\subsection{Outage Probability of Different Tiers and Types}\label{subsection_numerical_m}
In this subsection, we study the outage probabilities of different tiers and types. 
Fig.~\ref{FigA}(f) shows the analytical and simulation outage probabilities of different types and tiers. The simulation results agree with the analytical results. 
At tier-1, because $P_2>P_1$, the outage probability of type-$2$ tier-1 UEs is smaller. At tier-2, $Q_{12}>Q_{11}$ leads to smaller outage probability for type-2 tier-2 UEs; while $Q_{22}=Q_{21}$ leads to the same outage probabilities. Given an arbitrary typical UE, the Palm distribution of other UEs (i.e., interferers) remains the same as their original Poisson distribution. Thus, the distribution of interference remains the same. The effect of multiple types of UEs only manifests in the shape of the common distribution of interference.

\subsection{Exclusion Regions}\label{subsection_numerical_ex}
In this subsection, we discuss the outage performance under the influence of the two types of exclusion regions. 
Each simulation data point is averaged over $100000$ trials.

Under different radii of the exclusion regions, we study the outage probabilities at both tiers derived by our model and simulations, as shown in Figs.~\ref{FigA}(g)-\ref{FigB}(a).
These figures show that almost all value points computed by the model are within the $95\%$ confidence intervals, illustrating the correctness of our modeling of  two types of exclusion regions.

At tier-1 cells, the results show that the outage probability  does not decrease significantly when we increase the radius of exclusion regions (e.g., the outage probability is lowered by $0.015-0.03$ when we increase the radius from $100$ m to $500$ m). This is because, first, the interference at tier-1 BSs is not dominated by the interference from tier-2 UEs; and second, the exclusion regions are small, so that the probability that there are some tier-2 UEs in the exclusion regions causing large interference is small.  As a result, the outage probability  only slightly decreases as we introduce the exclusion regions.

Under the same radius, using UE-exclusion regions is more efficient to decrease the outage probability at tier-1 cells. This is because tier-2 UEs are strictly forbidden within a distance of $R_{e,2}$ from a tier-1 BS when UE exclusion is applied, while they may be located as near as $\left(R_{e,1}- R_1\right)^{+}$ under BS exclusion.

At tier-2 cells, the outage probability is influenced by the exclusion regions in more complicated ways: (a) Increasing $R_{e,1}$ or $R_{e,2}$ leads to less tier-2 UEs, which then deceases the outage probability. (b) Increasing $R_{e,1}$ or $R_{e,2}$ leads to a higher probability that a tier-2 cell is located at the edge of a tier-1 cell, where tier-1 UEs transmit at higher power levels, which in turn increases the outage probability.
(c) For the UE exclusion,  increasing $R_{e,2}$ leads to a higher probability that a tier-2 cell is overlapping with the exclusion regions, where tier-2 UEs are restricted, decreasing the intra-cell interference and outage probability.
Considering all of these factors,  if BS exclusion is applied, it is not obvious whether factor (a) or (b) dominates the other. Figs.~\ref{FigA}(i)-\ref{FigB}(a) indicate that the outage probability at tier-2 cells increases when $T=0.1$ and decreases when $T=1$, if we increase the radius of BS-exclusion regions.
On the other hand, if UE exclusion is applied, factor (c) dominates and thus the outage probability decreases when we increase the radius of exclusion regions.

\subsection{Optimal Targeted Received Power}\label{subsection_numerical_optimalP}
In this subsection, we present a numerical study on the optimal targeted received power ratio, as discussed in Section \ref{subsection_optimalP}. 
 Fig.~\ref{FigB}(b) shows the results.
Increasing $\frac{Q_{11}}{P_1}$ leads to higher outage probability for tier-1 UEs but lower outage probability for tier-2 UEs. The overall outage probability decreases and then increases, which is minimized around  $\frac{Q_{11}}{P_1}=15$ dB.

\subsection{Effects of Tier-2 UE Intensity Function}\label{subsection_numerical_differentV}
In this subsection, we discuss the effect of the tier-2 UE intensity function.
We focus on three intensity functions of tier-2 UEs, all in units/km$^2$: (a) $\nu_{11}(\bx)=20$, where UEs are homogeneously distributed; (b) $\nu_{11}^{+}(\bx)=30\frac{|\bx|}{R_1}$, where UEs are likely to locate at cell edges; and (c) $\nu_{11}^{-}(\bx)=60\frac{R_1-|\bx|}{R_1}$, where UEs are likely to locate near cell centers. Note that the tier-2 UE intensities are $0$ when $|\bx|>R_1$. In addition, the average numbers of tier-2 UEs in one tier-2 cell are the same in all of the three cases.

As expected in the discussion in Section \ref{subsection_differentV},  Figs.~\ref{FigB}(c)-\ref{FigB}(d) show that compared with $\nu_{11}(\bx)$,  $\nu_{11}^{+}(\bx)$ introduces higher interference from tier-2 UEs, causing higher outage probabilities at both tiers, while   $\nu_{11}^{-}(\bx)$ brings lower interference from tier-2 UEs, causing lower outage probabilities at both tiers.

\subsection{Orthogonal Multiple Access}\label{subsection_numerical_OFDM}
In this subsection, we present a numerical study on the systems using orthogonal multiple access schemes.  We simulate dependent thinning of UEs, instead of the independent thinning model used in our analysis.

Fig. \ref{FigB}(e) shows the comparison of simulated outage probabilities and analytical ones (derived through our modified model discussed in Section \ref{subsection_OFDM}). The number of resource blocks at each BS is $n=16$.
The figure shows that the analytical outage probabilities derived in Section \ref{subsection_OFDM} are only slightly smaller than simulated ones, suggesting that the modified version of our model provides useful approximations for systems using orthogonal multiple access schemes.

\subsection{Linear Intensity Tradeoff of Tier-2 cells}\label{subsection_numerical_intensityplan}

In this subsection, we investigate the linear intensity tradeoff  discussed in Section \ref{subsection_intensityplan}.
The outage probability constraints are $\mathbb{P}_{ target,1}=\mathbb{P}_{ target,1}=\mathbb{P}'_{ target,11}=\mathbb{P}'_{ target,21}=0.2$. All the other parameters are shown in Table \ref{tabel_para}. Computed by (\ref{formula_constraint2}) and (\ref{formula_constraint22}), the tradeoff between $\mu_1$ and $\mu_2$ follows the linear inequality,
\begin{equation}\label{formula_experiment}
0.0640\mu_1+0.0569\mu_2\leq 0.1761.
\end{equation}
 Note that (\ref{formula_experiment}) corresponds to the outage constraint  of type-1 tier-1 UEs, which dominates all the other outage constraints.

We also use simulation to search for the maximum $\mu_2$ under different $\mu_1$ values. Fig.~\ref{FigB}(f) shows the analytical and simulation results of the intensity tradeoff between $\mu_1$ and $\mu_2$.  The results show that the simulation results agree with the analytically obtained linear tradeoff in Section \ref{subsection_intensityplan}.

\subsection{Impact of Correlated Tier-1 UE or Tier-2 BS Locations}\label{subsection_numerical_correlated}

In this subsection, we study the performance under correlated tier-1 UE or tier-2 BS locations via simulation, in order to show that our model remains useful as an approximation when the spatial patterns are not strictly Poisson.

The correlated locations are generated as follows: In each trail of simulation, let $\mathbf{X}=(\bx_1,\bx_2,\ldots \bx_n)^\mathbf{T}$ denote original points corresponding to one type of tier-1 UEs or tier-2 BSs generated as PPP.
We set $\mathbf{X}'=\mathbf{K}^\mathbf{T}\mathbf{X}$ as the new coordinates by introducing correlations among the original coordinates. Then, $\mathbf{L}=\mathbf{K^T}\mathbf{K}$ is the covariance matrix of the new coordinates. We further set the $i$th row, $j$th column of $\mathbf{L}$,  $L_{ij}=\exp(-\alpha |i-j|^2)$, where  $\alpha$ is referred to as the correlation parameter. Smaller $\alpha$ indicates stronger correlations among the  points. Details to derive $\mathbf{K}$ from $\mathbf{L}$ can be found in Section 6.6 of \cite{basic}. Except the above location transformation, all the other parts of simulation remain the same.

Figs.~\ref{FigB}(g)-\ref{FigB}(h) show the outage probability of tier-1 UEs and tier-2 UEs under different correlation parameters.  Both figures show that the correlations of tier-1 UEs or tier-2 BSs will cause higher outage probabilities at both tiers. When the tier-1 UEs or tier-2 BSs are weakly correlated (e.g., $\alpha\geq 10^{2}$), the  gap between the correlated case and the non-correlated case is almost $0$. Even when the tier-2 BSs are strongly correlated (e.g., $\alpha\leq 10^{1.5}$), our model is still useful in approximating the outage probabilities.

\subsection{Impact of Correlated Tier-1 UE or Tier-2 BS Locations on the Intensity Planning Problem}\label{subsection_numerical_correlated2}
In this subsection, we study the intensity planning problem (presented in Section \ref{subsection_intensityplan}) under correlated  tier-1 UE or tier-2 BS locations. The correlated locations of tier-1 UEs or tier-2 BSs are generated by the method presented in Section \ref{subsection_numerical_correlated}.

 The network parameters are shown in Table \ref{tabel_para}. We also set the outage constraints for tier-1 UEs are $\mathbb{P}_{target,1}=\mathbb{P}_{target,2}=0.1$; and there are no  outage constraints for tier-2 UEs.
 The utility functions are $U_1(\mu_1)=1.5\ln(1+10\mu_1)$ and $U_2(\mu_2)=\ln(1+10\mu_2)$, $U(\mu_1, \mu_2)=U_1(\mu_1)+U_2(\mu_2)$. Each simulation data point is averaged over 30000 trials.

First, without  location correlations, we derive the optimal $U^{*}$, $\mu_1^{*}$, and $\mu_2^{*}$ analytically based on the method in Section \ref{subsection_intensityplan}. Second, given $\alpha$, we obtain  simulated outage probabilities of tier-1 UEs, $P_{out,1}(\alpha)$ and $P_{out,2}(\alpha)$, under $\mu_1^{*}$ and $\mu_2^{*}$. $P_{out,1}(\alpha)$ and $P_{out,2}(\alpha)$ are referred to as the \emph{relaxed outage probabilities}. Note that if the operator intends to attain $U^{*}$ under the correlated case with parameter $\alpha$, the original outage probability constraints should be modified as the \emph{relaxed outage probability constraints} (i.e., $\mathbb{P}_{target,1}(\alpha)=\mathbb{P}_{out,1}(\alpha)$ and $\mathbb{P}_{target,2}(\alpha)=\mathbb{P}_{out,2}(\alpha)$). Third, under these relaxed outage probability constraints, we can again derive the optimal utility $U(\alpha)$ analytically,  without location correlations. $U(\alpha)$ is referred to as the \emph{ameliorated utility} when the correlations diminish. Let $\eta(\alpha)=\frac{U(\alpha)-U^{*}}{U(\alpha)}$ denote relative performance gap. We study $\eta(\alpha)$ against $\alpha$ to show the performance gap under different $\alpha$ values.

In the presented experiment, the optimal solution of the original non-correlation case is $\mu_1^{*}=0.5204$, $\mu_2^{*}=0.4703$, and $U^{*}=4.4787$. Fig.~\ref{FigB}(i) shows $\eta(\alpha)$ against $\alpha$ with location correlations of tier-1 UEs and tier-2 BSs, respectively. The results show that the performance gap is small when the locations of tier-1 UEs or tier-2 BSs are weakly correlated ($\alpha\geq 10^2$).

\section{Conlusions} \label{section_conclusion}
In this paper, we propose a stochastic geometric model to accurately quantify the uplink interference and outage performance
of two-tier cellular networks with diverse users and tier-2 cells.
By applying our SAS approach, we  derive numerical expressions for the Laplace transform of
interference at both tiers, avoiding the approximations
required in prior works, leading to accurate numerical calculation of the
outage probability.
Our model is also able to capture the impact of two types of exclusion regions, in which either tier-2 base stations or tier-2 users are restricted in order to avoid cross-tier interference. As an application example, an intensity planning problem is investigated, in which  the outage probability  constraints are converted to linear intensity tradeoff, facilitating efficient solutions.
Finally, numerical studies further demonstrate the correctness and usefulness of our analysis.


\bibliographystyle{IEEEbib}
\bibliography{baoweiSG}

\end{document}